\documentclass[onecolumn]{aa}
\usepackage[varg]{txfonts}
\usepackage{graphicx}
\usepackage{natbib}
\usepackage{color}
\bibpunct{(}{)}{;}{a}{}{,}
\DeclareMathOperator{\sgn}{sgn}
\DeclareMathOperator{\Heaviside}{H}
\usepackage{amstext}

\begin{document}


\title{Five methods for determining pattern speeds in galaxies}
\subtitle{I. Methods}

\author{Daniel Pfenniger\inst{\ref{inst1}}\and Kanak Saha\inst{\ref{inst2}}\and Yu-Ting Wu\inst{\ref{inst3}}}

\institute{Geneva Observatory, University of Geneva, CH-1290 Versoix, Switzerland\label{inst1}\and 
{Inter-University Centre for Astronomy and Astrophysics, Pune 411007, India}\label{inst2}\and
{National Astronomical Observatory of Japan, Mitaka, Tokyo 181-8588, Japan}\label{inst3}}

\date{Received -- -- 2022/ Accepted -- -- 2022}
\abstract 
{ After matter distribution and kinematics, the bar or spiral pattern speeds are the next fundamental parameters determining the
  dynamics of a galaxy.  }
{New or refined methods for determining instantaneous scalar and vector pattern speeds from a restricted domain are developed for
  applications in $N$-body simulations or in galaxies such as the Milky Way, for which the stellar coordinates become increasingly
  better known.}
{The general feature used throughout follows from the fact that the time derivative of a function of the coordinates is linearly
  proportional to its rotation rate and its particle velocities.  Knowing these therefore allows retrieving the instantaneous
  pattern speed vector by linear optimization.  Similarly, if an invariant function depends on the position and velocities, then its
  instantaneous rotation vectors in space can be retrieved. Knowing the accelerations also allows determining the pattern rotation
  of velocity space.  }
{ The first three methods are based on the assumed rotational invariance of functions at each point in space or velocity space: 1)
  The 6D invariant function method, measuring the pattern speed vectors in space and velocity space, 2) the differential/regional 3D
  Tremaine-Weinberg method, evaluated over regions with a high signal-to-noise ratio, 3) the 3D Jacobi integral method, yielding the
  potential pattern speed.  Extensions to derive the rotation center position, speed, and acceleration are introduced in the first
  and third methods.  The last two methods are based on the assumed invariance of average functions of the particle coordinates: 4)
  The 2D and 3D moment of inertia methods by using the derivative of the singular value decomposition, 5) the 2D Fourier method (3D
  for $m=2$ mode), giving the mode rotation speeds.  Pattern speed accelerations are also derived in the fourth and fifth methods.
}
{ Depending on the available data in specific problems, the different methods provide a choice of approaches.}

\keywords{Galaxy: fundamental parameters -- Galaxy:kinematics and dynamics
          -- Galaxies: fundamental parameters -- Galaxies:kinematics and dynamics 
          -- Methods:numerical -- Methods:data analysis}

\maketitle

\titlerunning{Five Methods for Determining Pattern Speeds }
\authorrunning{Pfenniger et al.}


\section{Introduction} 
\label{intro} 

The rotation frequency $\Omega_\textrm{p}$, also called the pattern speed, of morphological features such as bars or spiral arms is
a crucial parameter in most dynamical studies of disk galaxies because it allows describing the local dynamics in a rotating frame
where time dependence is reduced. Physical and numerical reality is more complicated, however, because most self-consistent galaxy
models, even when they do not include a dissipative gas component, show that several flexible patterns with distinct speeds coexist
\citep{SparkeSellwood87, Wozniak2015, Wozniak2020, WPT16, WPT18}. This implies that the system is not strictly time-invariant at
meso- and macro-scales in any rotating frame.  Therefore, periodic or secular angular speed variations must be expected, and the
amplitude of the time fluctuations is an aspect of galactic dynamics that needs to be better examined.  The constancy of spiral arm
pattern speeds has been a fundamental assumption in the density wave theory of \cite{LinShu64} all along, but this theory has been
challenged \citep{Sellwood11}: Spiral arms might be winding up, or be transient \citep{ToomreKalnajs91} in view of the efficient
swing amplification of small perturbations in cold disks \citep{JulianToomre66}.  Long-lived ($\sim 5$\,Gyr) spirals are not
excluded \citep{SahaElmegreen2016} in particular conditions, however.  More quantitative numerical investigations of these questions
in fully consistent models are therefore required.  In the meanwhile, evaluations of the instantaneous pattern speeds of spatially
limited features are required in order to be able to verify whether long-time pattern speed averages, as commonly calculated, make
sense for disk dynamics, and to quantitatively determine the errors when assuming a single global pattern speed.

For solid bodies, there is no internal streaming, thus the instantaneous rotation speed and instantaneous pattern speed are
identical.  More realistically, in stellar and fluid systems, the mean rotation of their matter and the rotation of one or several
patterns may all be different, and differ spatially, as was clearly demonstrated in \cite{WPT18}, because bars and spirals may be
flexible structures.

The precise definition of ``pattern speed'' is rarely specified, while a ``pattern'' may mean different things: it can be just a
special morphological feature, or the shape of a mass distribution in space, kinematical space, or phase-space, or in more abstract
spaces, such as frequency (Fourier) space.  To characterize a shape, the gradient of observable quantities, such as the mean mass
distribution, must be available.  The motion of this gradient allows us to follow the shape in time, and, if it rotates uniformly
around a center, it can describe the pattern rotation speed.  Assuming a single pattern rotation speed therefore assumes that all
gradients of the observable quantity over the examined region rotate rigidly, at the same angular speed, around a common center.

That a stellar or particle system displays a single pattern speed everywhere is very particular. Typical
galaxies or $N$-body models of galaxies may display an approximate constant pattern speed over only a subset of the whole. When a
pattern speed is estimated, it therefore is important to understand the possible bias introduced by selecting a specific particle
subset.  Unless a continuous differentiable function is involved, in which case the motion of the function gradient at any point
yields a speed, a pattern in galaxies or in simulations is a statistical property over many particles. Choosing a specific subset
may therefore introduce a selection bias.  For example, the pattern speed of a single particle can only be its speed.  The
pattern speed of a well-grouped set of particles (a cluster) is its average speed.  The pattern speed of a ring of particles is not
the average circular speed of its particles, but the speed of a density gradient moving as a single wave. Depending on the subset
selection, the speed of a compact subset of stars may therefore reflect the local streaming motion of this subset more than the
motion of a pattern of larger size.

In numerical work, most often, in $N$-body simulations, full information about the particle distribution and properties is
available, so that the assumptions made for galaxies can in principle be checked in detail. A popular approach to determine an
average pattern speed is to Fourier transform, over angle and time, the particle distribution in radial bins
\citep{SellwoodAthanassoula86,MassetTagger97,RautiainenSalo99}. Because the simulations are not time periodic, a finite time window
is used that is longer than the pattern period, but shorter than the simulation length. Power peaks of a given Fourier mode in the
time domain at a given radius reveal the average mode pattern speed(s) over the chosen time window. The power spectrum indicates
whether a single or multiple distinct patterns coexist on average at any given radius. The sampling of shortest and longest
time-intervals bounds the examined average frequencies, but also introduce aliases. Strictly, any power spectrum different from a
single sharply peaked function (the width of which should be consistent with the particle noise) then indicates time variability.

A direct straight method for finding the almost instantaneous bar pattern speed has been used, for instance, by
\cite{Pfenniger_Friedli91}. The method consists of measuring the angle of its inertia tensor major axis separated by a short time
interval, where the inertia tensor is calculated with the particles within a cylinder as large as the bar. This method provides an
almost instantaneous pattern speed averaged over a time step, but the result may depend on the adopted cylinder size.  Its drawback
is that it requires two particle snapshots, which is unnecessary, as we show in Sect.~\ref{sec:MOMENT}.

The current most popular method to find the instantaneous $\Omega_\textrm{p}$ has been proposed by \cite{TremaineWeinberg84} (TW),
with some modifications in \cite{Merrifield_etal06} and \cite{Meidt_etal08}, where radial variations are considered.  The
conservation of the observed population projected in the galactic plane is assumed. Integrating spatial gradients over lines allows
recovering the bar pattern speed when only the line-of-sight velocity component and the disk inclination angle are known. The
problem with the TW method is that the pattern speed is assumed to be uniform over the whole observed disk, which is questionable in
view of the known pattern speed differences between the bar and the spiral arms in $N$-body simulations
\citep{SparkeSellwood87,RautiainenSalo99}. The possible interaction between the bars and spiral arms might also affect the pattern
speed measurement \citep{Hilmietal2020}.  Furthermore, nested double bars are common \citep{Peter2004}, and are expected to rotate
at very distinct speeds \citep{Pfenniger_Norman_90}, and was indeed found in simulations \citep[e.g.,][]{FriedliMartinet1993,
  DebattistaShen2007,SahaMaciejewski2013,WPT16,WPT18}.  A more flexible method giving the instantaneous pattern speed over a spatial
domain relevant for the addressed problem appears to be required.

Methods based on the kinematic signature of specific resonances, such as the Milky Way bar outer Lindblad resonance observed in a
star sample around the Sun (e.g., \cite{Monari_etal17}), are global and implicitly assume that the disk dynamics at the Sun is not
strongly perturbed by the local spiral arms, which rotate possibly with a distinct pattern speed, as is frequently observed in
$N$-body simulations.  The mere presence of self-gravitating spiral arms around the Sun suggests the opposite, however:
Local forces are substantially dominated by the influence of local disk matter.

Offset bars, now seen in more than a dozen galaxies \citep{deSwardtetal2015,Kruketal2017}, present another kind of problem. Because
the offset bar may not necessarily be at the kinematical center of the host galaxy or at the geometrical center of the outer disk,
the bar pattern, including its central density peak, may rotate both about itself and as a whole about the galaxy center, performing
a kind of epicyclic motion around the center of mass, a superposition of two or more rotational motions.

Here we concentrate on methods providing a single instantaneous pattern speed and possibly local pattern speeds in $N$-body models
because the information contained in a particle snapshot should be sufficient to find them in principle, and, in some cases, their
pattern acceleration as well. In case of multiple rotational components, to first order, we search only for the fastest rotational
component in the system and its instantaneous rotation center.  The slower additional rotational components are then approximated to
first order with a constant translation of the rotation center.  Proceeding beyond first order may be vain as the pattern speed
invariance hypothesis is likely to be invalid over longer time intervals.

Some of these methods may hopefully be useful for interpreting astrometric data of the Milky Way.  Precise astrometric surveys, such
as the Gaia survey \citep{Gaia2018}, start to provide full phase-space coordinate information for large Milky Way star samples.
Alternative methods to determine a pattern speed, in particular of the Milky Way bar, for which a number of methods and tracers
already exist \citep[e.g.,][]{Debattistaetal2002,Gerhard2011, Minchevetal2007, Sandersetal2019, ClarkeGerhard2022}, may be found to
be useful and complementary.

It might be thought that any method should give the same pattern speed value. Using three of methods exposed here, \cite{WPT18}
showed that according to the adopted shape definition, somewhat different pattern speeds are found in the same model because
different methods measure different characteristics of the not strictly rigidly and constantly rotating pattern.  This is especially
relevant for the double-bar galaxy models used in \cite{WPT18}, or for non-rigidly rotating spiral arms.  In regions in which the
system is strongly time dependent with respect to the local dynamical time, for example, at the interface between two nested bars or
between a bar and the surrounding spiral arms around the corotation radius, methods based on the assumption of a constant pattern
speed obviously break down, as shown in \cite{WPT16,WPT18}, because the single steady pattern assumption is incorrect.

In Sect.\ \ref{sec:FEATURE} we discuss a naive method along with its shortcomings.
In Sect.~\ref{sec:ivfunc} we derive the scalar or vector pattern speed in space and velocity space from any observable function of
space and/or velocity space that is assumed invariant by rotation and translation, with a possibly unknown rotation center.
In Sect.\ \ref{sec:LTW}, as a generalization of the TW method, we examine the local and regional TW methods (herafter LTW and RTW),
and describe the criterion where in observations, the derived pattern speed is a priori the least affected by measurement errors.
Then, in Sect.\ \ref{sec:JACOBI}, we use the Jacobi integral, or equivalently, the rotational invariance of the mean potential, to
determine the potential pattern speed.
In Sect.\ \ref{sec:MOMENT} we consider how second-moment tensors in ordinary space, or velocity space, allow us to determine
$\Omega_\textrm{p}$, when the rotation axis direction is known or not.
In Sect.\ \ref{sec:FOURIER} we use the Fourier space phase information to analyze how circular wave modes rotate, which is for mode
2 a similar but distinct approach from the moment methods.
Finally, we conclude in Sect.\ \ref{sec:CONCLUSIONS}.


\section{Morphological feature method}\label{sec:FEATURE}
A direct method (not counted in the five methods) is to use a clear morphological feature, assumed to be corotating with the
figure. For example, a barred spiral can display a straight bar and a symmetric two-arm spiral apparently starting at the ends of
the bar.  The intersection points of a logarithmic spiral fitted along the spiral arms and a straight line along the bar major axis
define two symmetric points that can be assumed to corotate with the bar. Dynamically, the most obvious explanation for this feature
is that these points represent the stationary Lagrange points $L_{1,2}$ of the bar pattern \citep{Pfenniger1990}.  Other examples in
the literature attempted to use the Lagrange points $L_{4,5}$ \citep[e.g.,][]{Shimizu2012}, the location of which can be inferred
from various tracers, such as a star-forming region or a local minimum of surface brightness in case of unstable $L_{4,5}$ for
strong bars.

When we know both the position $(x,y)$ and the velocity in an inertial frame $(v_x,v_y)$ of one of these points in the galaxy plane,
the angle is found by
\begin{equation}
  \phi = \arctan(y(t),x(t)), 
\end{equation}
and the pattern speed is its time derivative,
\begin{equation}
  \Omega \equiv {d\over dt}\arctan(y(t),x(t)) = {x v_y - y v_x \over x^2+y^2} = {v_t \over R}, 
\end{equation}
where $v_t$ is the tangential velocity component of the point $(x,y)$, and $R = (x^2+y^2)^{1/2}$ is its distance to origin.

This method clearly relies on the assumption that the chosen point is indeed stationary in some rotating frame.  For $N$-body bars,
\cite{WPT16} have shown that the instantaneous equilibrium points between a bar and spiral arms may oscillate with a substantial
amplitude in any rotating frame, including the bar frame. No strict Lagrange points therefore exist because of the perturbation by
the adjacent spiral pattern that rotates at a different speed.  The assumptions made about the pattern that we wish to measure are
therefore clearly crucial. In some cases, no strictly constant pattern speed even exists.
  
 
\section{Invariant function method}\label{sec:ivfunc}

A pattern can be defined by the mean of a function $f(\vec x, \vec v, t)$ (e.g., the phase-space distribution function of stars, the
mass density of a particular stellar population, or the mean gravitational potential). This function is assumed to be invariant
either 1) after an infinitesimal time $dt$, or 2) at fixed time $t$ after an infinitesimal rigid rotation $dt\,\vec \Omega $ about
some rotation center in configuration space $\{\vec x\}$, and likewise an infinitesimal rigid rotation $dt\,\vec \omega $ about some
rotation center in velocity space $\{\vec v\}$.

\subsection{Rotation about the origin} 

First, we start with the simplest case. The rotation center is assumed to be fixed and located at the origin.  We generalize the
derivation used by \cite{TremaineWeinberg84} in their pioneer paper to more cases.  The function $f(\vec x, \vec v, t)$ invariance,
either after an infinitesimal time displacement or after an infinitesimal solid rotation about the origin, reads
\begin{equation}\label{equ:infrot}
  f(\vec x,\vec v, t+dt) = f(\vec x + dt\, \vec \Omega \wedge \vec x, \vec v + dt\, \vec \omega \wedge \vec v ,t).
\end{equation}
A priori, the pattern speeds $\vec \Omega$ and $\vec \omega$ do not need to be equal.  For the sake of rigor, we verify in Appendix
B the mere existence of such functions by displaying an explicit class of these.

Expanding both sides of the equation to first order in time leads to a linear equation for $\vec \Omega$ and $\vec \omega$ that can
be reformulated more conveniently by using the dot and cross products property
$\vec a \cdot (\vec b\wedge \vec c) = (\vec c \wedge \vec a) \cdot \vec b$,
\begin{eqnarray}\label{equ:infrot2}
 \partial_tf  ~=~ 
      \partial_{\vec x}f \cdot(\vec\Omega \wedge \vec x) + 
      \partial_{\vec v}f \cdot(\vec\omega \wedge \vec v) ~ = ~ 
      \vec x \wedge \partial_{\vec x}f \cdot\vec\Omega + \vec v \wedge \partial_{\vec v}f \cdot\vec\omega .
\end{eqnarray}

If the direction unit vector $\vec n$ of the rotation axis is known, $\vec \Omega = \Omega \vec n $, and the two frequencies are the
same, $\vec \omega = \vec \Omega$, we can then attempt to solve this linear equation for $\Omega$ provided the partial derivatives
of $f$ are known at a given phase space point $(\vec x, \vec v)$ at time $t$.  Rotating the axes such that $\vec n = (0,0,1)$, we
find
\begin{eqnarray}
  \Omega  ~=~ { \partial_tf \over x \partial_y f -  y \partial_x f + v_x \partial_{v_y} -  v_y \partial_{v_x} f}.
\end{eqnarray}

If $\vec n$ is not known and $\vec \omega = \vec \Omega $, we need $n\geq 3$ independent positions $(\vec x_i, \vec v_i)$,
$i=1,2,\ldots n$ to solve by linear optimization, such as the least-squares method, the linear system of equations
\begin{equation} \label{eq:LSf1}
  \underbrace{
  \left( 
    \begin{array}{c}
      \vec x_1 \wedge  \partial_{\vec x}f_1 + \vec v_1 \wedge  \partial_{\vec v}f_1 \\
       \vdots \\
      \vec x_n \wedge  \partial_{\vec x}f_n + \vec v_n \wedge  \partial_{\vec v}f_n \\
    \end{array} 
  \right) 
  }_{n\times 3 \ \textrm{\scriptsize matrix}}
\underbrace{
  \vec\Omega
 }_{3\times 1 \ \textrm{\scriptsize vector}}
   \approx
  \underbrace{
  \left(
    \begin{array}{c}
      \partial_tf_1  \\
      \vdots   \\
      \partial_tf_n
    \end{array}
  \right)
  }_{n\times 1 \ \textrm{\scriptsize vector}}
 ,
\end{equation}
where $f_i$ means $f(\vec x_i, \vec v_i, t)$. In the presence of noise, no reliable solution exists if the $n\times 3$ left-hand
side matrix is degenerate, however.

The least-squares solution of a linear system $Ax\approx b$, where $A$ is the $n\times m$ matrix ($n>m$), and $b$ the $n\times 1$
vector, is the unique solution for the $m\time 1$ vector $x$ minimizing the Euclidian $L_2$ norm $||Ax-b||_2$, and, when $A$ is
degenerate, the minimum $||x||_2$.  In addition to the traditional least-squares method, methods that minimize other norms are
available, such as the $L_1$ norm, which may be less sensitive to outlier data points \citep{BoydVandenberghe2004}. In contrast to
the $L_2$ norm, however, a $L_1$ norm solution may not always exist.

If $\vec \omega \neq \vec \Omega $, we need $n\geq 6$ independent positions $(\vec x_i, \vec v_i)$, $i=1,2,\ldots n$ to solve the
linear system as above,
\begin{equation}\label{eq:LSf2}
  \underbrace{
  \left( 
    \begin{array}{cc}
      \vec x_1 \wedge  \partial_{\vec x}f_1 &  \vec v_1 \wedge  \partial_{\vec v}f_1 \\
       \vdots & \vdots \\
      \vec x_n \wedge  \partial_{\vec x}f_n &  \vec v_n \wedge  \partial_{\vec v}f_n \\
    \end{array} 
  \right) 
  }_{n\times 6 \ \textrm{\scriptsize matrix}}
\underbrace{
  \left(
     \begin{array}{c}
       \vec\Omega \\
       \vec\omega
     \end{array}
  \right)
 }_{6\times 1 \ \textrm{\scriptsize vector}}
   \approx
  \underbrace{
  \left(
    \begin{array}{c}
      \partial_tf_1  \\
      \vdots   \\
      \partial_tf_n
    \end{array}
  \right)
  }_{n\times 1 \ \textrm{\scriptsize vector}}
 .
\end{equation}

In these linear optimization problems, each line of Eqs.(\ref{eq:LSf1}--\ref{eq:LSf2}) can always be multiplied left and right by a
weight $w_i$ to weight the data points differently. For example, the weight can be inversely proportional to the error $\sigma_i$
expected on the $i\text{th}^\mathrm{}$ data point, $w_i = \sigma_i^{-1}$. For the $L_2$ norm, this weight is equivalent to
performing a $\chi^2$ minimization.

\subsection{Incompressible function $f$} 

If $f$ conserves phase-space density, such as for Boltzmann's collisionless equation, where $f(\vec x,\vec v, t)$ is the phase-space
distribution function,
\begin{equation}
\partial_t f + \vec v \cdot \partial_{\vec x}f + \vec a \cdot \partial_{\vec v} f=0 , 
\end{equation}
where $\vec a \equiv \dot{\vec v}$ is the acceleration, then $\partial_t f$ can be eliminated. 

When $\vec \omega = \vec \Omega $, we obtain 
\begin{equation}\label{eq:omegainvfunc}
  \left( 
    \begin{array}{c}
      \vec x_1 \wedge  \partial_{\vec x}f_1 + \vec v_1 \wedge  \partial_{\vec v}f_1 \\
       \vdots\\
      \vec x_n \wedge  \partial_{\vec x}f_n + \vec v_n \wedge  \partial_{\vec v}f_n
    \end{array} 
  \right) 
  \vec\Omega
  \approx
  -\left(
    \begin{array}{c}
      \vec v_1 \cdot \partial_{\vec x}f_1 + \vec a_1 \cdot \partial_{\vec v}f_1  \\
      \vdots \\
      \vec v_n \cdot \partial_{\vec x}f_n + \vec a_n \cdot \partial_{\vec v}f_n
    \end{array}
  \right) , 
\end{equation}
and when $\vec \omega \neq \vec \Omega $,
\begin{equation}\label{eq:2omegainvfunc}
  \left( 
    \begin{array}{cc}
      \vec x_1 \wedge  \partial_{\vec x}f_1 & \vec v_1 \wedge  \partial_{\vec v}f_1 \\
       \vdots & \vdots \\
      \vec x_n \wedge  \partial_{\vec x}f_n & \vec v_n \wedge  \partial_{\vec v}f_n
    \end{array} 
    \right)
    \left(
     \begin{array}{c}
       \vec\Omega \\
       \vec\omega
     \end{array}
  \right)
  \approx
  -\left(
    \begin{array}{c}
      \vec v_1 \cdot \partial_{\vec x}f_1 + \vec a_1 \cdot \partial_{\vec v}f_1  \\
      \vdots \\
      \vec v_n \cdot \partial_{\vec x}f_n + \vec a_n \cdot \partial_{\vec v}f_n
    \end{array}
  \right) , 
\end{equation}

Thus, when $f$ depends on $\vec v$, we need to also know the acceleration (or force) field $\vec a$, which is typically minus the
gradient of the gravitational potential $\Phi(\vec x)$.  When $f$ does not depend on $\vec v$, all the $\partial_{\vec v}f_i$ terms
cancel, $\vec a$ is not needed, and as it should, $\vec \omega$ then remains undetermined.

\subsection{Mass-conserving function $f(\vec x, t)$} 

Likewise, when $f$ conserves mass in space, for instance, the mass density in a fluid, often called $\rho$, and
$\vec u = \vec u(\vec x, t)$ is the velocity field, $\partial_t f$ can be eliminated by the continuity equation
\begin{equation}
\partial_t f + \partial_{\vec x}\cdot ( f \vec u) =0 . 
\end{equation}
Because $\vec u$ is a function of $\vec x$, it rotates in space at the same speed as in velocity space, so that $\Omega = \omega$.
Determining $\vec \Omega$ then requires knowing only the spatial gradients of $f$ and $\vec u$ at time $t$.

\subsection{Translation and rotation about a known center}
When $f$ is invariant only when it rotates about the center at frequencies $\vec \Omega$ and $\vec \omega$, and when in addition,
the center moves at velocity $\vec V$ and acceleration $\vec A$, Eq.\ (\ref{equ:infrot}) is modified as
\begin{equation}\label{equ:infrottrans}
  f(\vec x,\vec v, t \!+ \!dt) = f(\vec x + dt\,(\vec V \!+\! \vec \Omega \wedge \vec x), \vec v + dt\, (\vec A \!+\! \vec \omega
  \wedge \vec v) ,t).
\end{equation}
Expanding both sides of the equation as before to first order in time leads to a linear equation system for $\vec V$, $\vec A$,
$\vec \Omega$, and $\vec \omega$,
\begin{equation}\label{eq:omegaVA}
  \underbrace{
  \left( 
    \begin{array}{cccc}
      \vec x_1  \wedge  \partial_{\vec x}f_1 & \vec v_1  \wedge  \partial_{\vec v}f_1 & \partial_{\vec x}f_1 & \partial_{\vec v}f_1\\
       \vdots & \vdots & \vdots & \vdots \\
      \vec x_n  \wedge  \partial_{\vec x}f_n & \vec v_n  \wedge  \partial_{\vec v}f_n & \partial_{\vec x}f_n & \partial_{\vec v}f_n\\
    \end{array} 
  \right) 
  }_{n\times 12 \ \textrm{\scriptsize matrix}}
  \underbrace{
  \left( 
  \begin{array}{c}
     \vec\Omega \\ 
     \vec\omega \\ 
     \vec V\\
     \vec A\\
  \end{array}
  \right)
  }_{12\times 1 \ \textrm{\scriptsize vector}}
   \approx
  \underbrace{
  \left(
    \begin{array}{c}
      \partial_tf_1  \\
      \vdots   \\
      \partial_tf_n
    \end{array}
  \right)
  }_{n\times 1 \ \textrm{\scriptsize vector}}
 .
\end{equation}
As above, we can replace the $\partial_tf_i$ terms if $f$ follows a conservation law.  In principle, we can therefore quantify by
how much the system center of rotation moves or accelerates with the same data required to solve Eq.\ (\ref{eq:2omegainvfunc}) by
solving Eq.\ (\ref{eq:omegaVA}) for $\vec \Omega$, $\vec V$, and $\vec A$ with at least $n\geq 12$ data points by linear
optimization.

Because Eq.(\ref{eq:omegaVA}) is dimensionally inhomogeneous, it may be useful to scale the left-hand side matrix columns and vector
lines by appropriate constants.  For example, when we use length and velocity scale factors $R_0$ and $V_0$, the system becomes
\begin{equation}\label{eq:omegaVAb}
  \left( 
    \begin{array}{cccc}
      \vec x_1  \wedge  \partial_{\vec x}f_1 & \vec v_1  \wedge  \partial_{\vec v}f_1 & R_0\partial_{\vec x}f_1 & V_0\partial_{\vec v}f_1\\
       \vdots & \vdots & \vdots & \vdots \\
      \vec x_n  \wedge  \partial_{\vec x}f_n & \vec v_n  \wedge  \partial_{\vec v}f_n & R_0\partial_{\vec x}f_n & V_0\partial_{\vec v}f_n\\
    \end{array} 
  \right) 
  \left( 
  \begin{array}{c}
     \vec \Omega \\ 
     \vec \omega \\ 
     \vec \Upsilon \\
     \vec \Lambda
  \end{array}
  \right)
 \approx
  \left(
    \begin{array}{c}
      \partial_tf_1  \\
      \vdots   \\
      \partial_tf_n
    \end{array}
  \right)
  , 
\end{equation}
and $\vec V = R_0\vec \Upsilon$, $\vec A = V_0\vec \Lambda$.
 
\subsection{Translation and rotation about an unknown center}
When the center of rotation is unknown both in space and velocity space, the true centers
of rotation would be $\vec x_0$ and $\vec v_0$ , respectively.  The invariant $f$ would follow the following constraint to first order
in $dt$,%
\begin{equation}\label{equ:infrottranscenter}
  f(\vec x,\vec v, t \!+ \!dt) = f(\vec x + dt\,(\vec V \!+\! \vec \Omega \wedge (\vec x-\vec x_0)), \vec v + dt\, (\vec A \!+\! \vec \omega
  \wedge (\vec v-\vec v_0) ) ,t).
\end{equation}
Expanding both sides of the equation as before to first order in time leads to a quadratic (hyperbolic) equation system for
$\vec x_0$, $\vec v_0$, $\vec \Omega$, and $\vec \omega$, and linear for $\vec V, \vec A$,
\begin{small}
\begin{eqnarray}\label{eq:omegaVAc}
\underbrace{
  \left( 
    \begin{array}{cccc}
      (\vec x_1-\vec x_0)  \wedge  \partial_{\vec x}f_1 & (\vec v_1-\vec v_0)  \wedge  \partial_{\vec v}f_1 & \partial_{\vec x}f_1 & \partial_{\vec v}f_1 \\
       \vdots & \vdots & \vdots & \vdots \\
      (\vec x_n-\vec x_0)  \wedge  \partial_{\vec x}f_n & (\vec v_n-\vec v_0)  \wedge  \partial_{\vec v}f_n & \partial_{\vec x}f_n & \partial_{\vec v}f_n \\
    \end{array} 
  \right) 
  }_{n\times 12 \ \textrm{\scriptsize matrix}}
  \underbrace{
  \left( 
  \begin{array}{c}
     \vec\Omega \\ 
     \vec\omega \\ 
     \vec V\\
     \vec A\\
   \end{array}
  \right)
  }_{12\times 1 \ \textrm{\scriptsize vector}}
  & \approx  &
  \underbrace{
  \left(
    \begin{array}{c}
      \partial_tf_1  \\
      \vdots   \\
      \partial_tf_n
    \end{array}
  \right)
  }_{n\times 1 \ \textrm{\scriptsize vector}}
  .
\end{eqnarray}
Rearranging the terms, we can separate the linear and quadratic parts,
\begin{eqnarray}\label{eq:omegaVAc2}
 \underbrace{
  \left( 
    \begin{array}{cc}
      \vec \Omega  \wedge  \partial_{\vec x}f_1 & \vec \omega  \wedge  \partial_{\vec v}f_1  \\
       \vdots & \vdots  \\
      \vec \Omega  \wedge  \partial_{\vec x}f_n & \vec \omega  \wedge  \partial_{\vec v}f_n  \\
    \end{array} 
  \right) 
  }_{n\times 6 \ \textrm{\scriptsize matrix}}
  \underbrace{
  \left( 
  \begin{array}{c}
     \vec x_0 \\ 
     \vec v_0 \\ 
   \end{array}
  \right)
  }_{6\times 1 \ \textrm{\scriptsize vector}}
  & \approx  &
  \underbrace{
  \left(
    \begin{array}{c}
      \partial_tf_1  \\
      \vdots   \\
      \partial_tf_n
    \end{array}
  \right)
  }_{n\times 1 \ \textrm{\scriptsize vector}}
  -
  \underbrace{
  \left( 
    \begin{array}{cccc}
      \vec x_1  \wedge  \partial_{\vec x}f_1 & \vec v_1  \wedge  \partial_{\vec v}f_1 & \partial_{\vec x}f_1 & \partial_{\vec v}f_1 \\
       \vdots & \vdots & \vdots & \vdots \\
      \vec x_n  \wedge  \partial_{\vec x}f_n & \vec v_n  \wedge  \partial_{\vec v}f_n & \partial_{\vec x}f_n & \partial_{\vec v}f_n \\
    \end{array} 
  \right) 
  }_{n\times 12 \ \textrm{\scriptsize matrix}}
  \underbrace{
  \left( 
  \begin{array}{c}
     \vec\Omega \\ 
     \vec\omega \\ 
     \vec V\\
     \vec A\\
   \end{array}
  \right)
  }_{12\times 1 \ \textrm{\scriptsize vector}}
  .
\end{eqnarray}
\end{small}
An iterative approach to determine a least-squares solution for this system might be to start by guessing initial values for
$\vec x_0$, $\vec v_0$, and then to solve Eq.~(\ref{eq:omegaVAc}) for $\vec\Omega$, $\vec\omega$, $\vec V$, and $\vec A$ by linear
least-squares.  Using these values, we could then solve for $\vec x_0$, $\vec v_0$ Eq.~(\ref{eq:omegaVAc2}) by linear least-squares.
Iterating this process a number of times might converge to a solution.  Otherwise, quadratic or nonlinear least-squares algorithms
could be required.  In general, we expect either no, one, or several real solutions with quadratic problems.

\section{Local and regional 3D Tremaine-Weinberg methods (LTW and RTW methods)} \label{sec:LTW} 

The TW method \citep{TremaineWeinberg84} is based on a particular case of invariant function, as discussed in Sect.\
\ref{sec:ivfunc}, where the surface density
\begin{equation}
\Sigma(x,y,t) = \int\! dz \int \!d^3v \, f(x,y,z,\vec v, t)
\end{equation}
is a projection of a thin distribution function $f$ in the galactic plane of a tracer assumed to verify the continuity equation
$\partial_t\Sigma + \partial_{\vec x} \cdot \left(\Sigma {\vec u}\right) = 0$, where ${\vec u}(x,y)$ is the local mass-weighted
average velocity in the plane $x-y$.  Because $\Sigma$ does not depend on $\vec v$, this method does not require  knowledge of the force
field.  In the original method, the direction of the rotation axis $\vec n$ is assumed to be known.  Using
Eq.~(\ref{eq:omegainvfunc}) at a single position, and integrating over all velocities,
\begin{equation} \label{eq:LTW}
\left(\vec x \wedge \partial_{\vec x}\Sigma \right)\cdot \vec n \ \Omega_\mathrm{LTW} = 
  - \partial_{\vec x}\cdot \left( \Sigma  {\vec u}\right).
\end{equation}
In principle, a single local calculation at $\vec x$ is therefore sufficient to find $\Omega_\mathrm{LTW}$. Because this calculation
involves gradients, however, observational errors may be an issue, as taking the gradient of data, for example, by finite
differences, typically amplifies errors.  Furthermore, even when precise gradients are available, the coefficient
$\vec x \wedge \partial_{\vec x}\Sigma$ in Eq.\ (\ref{eq:LTW}) may vanish (e.g., when the azimuthal density profile is an extremum,
or more generally, when it is tangent to a circle centered at the origin), so that large errors are contributed by all the regions
that are locally rotationally symmetric.  This occurs, for example, close to the main axes of a bar, but also in the disk.  By
weighting each data point line in Eq.(\ref{eq:LTW}) by a coefficient $w_i$ proportional to the error inverse, we can thus achieve a
$\chi^2$ minimization.

The original TW method attempts to minimize noise by using weighted averages of Eq.\ (\ref{eq:LTW}) along lines. In these line
integrals, the high-noise segments mostly contribute to increasing errors, the above-noise signal segments alone do contribute to
finding $\Omega$.  Extending the integration interval to infinity and performing integration by parts avoids formally dealing with
gradients in the integrals, but this does not remove the measurement noise, and it necessarily includes contributions outside the
relevant region and in this way, it may add systematic errors in case of multiple pattern speeds.  Since $N$-body bars and real
galaxies have different pattern speeds in their bar(s) and surrounding spiral arms \citep{SellwoodSparke88, WPT16, WPT18}, and
double-bar systems are frequent \citep{Peter2004}, it is therefore necessary to consider local or regional evaluations of pattern
speeds.

Local methods use data gradients, however, which, if done without attention, are prone to amplifying noise.  In the literature, many
approaches exist for evaluating gradients of noisy data. For the systems considered here, methods that smooth data locally have to
be preferred over methods that use global characteristics of noise.  One of the oldest local method is the Savitzky-Golay filter
\citep{SavitzkyGolay1964}. Spectral methods that smooth the gradient operator instead of smoothing data are intrinsically nonlocal.
In astrophysical particle systems, the smoothed-particle hydrodynamics technique \citep{GingoldMonhagan1977, Lucy1977} is frequently
used to evaluate the density gradients and other physical quantities attached to particles.  A smoothing kernel convolved spatially
over the particle data provides smooth density and velocity functions.  In numerical mathematics, more advanced methods have been
proposed that not only provide accurate gradients, but also preserve discontinuities \citep[e.g.,][]{Chartrand2011,Chartrand2017}.
We refer to \cite{VanBreugel2020} for a recent short review of the problem.  In view of the general need in science and engineering
to extract derivatives of data, the question will continue to progress in the following years.  Below we assume that the problem of
taking gradients of data has been sufficiently well mastered.

\subsection{Local method (LTW)}

An alternative approach to TW is to choose a number of selected locations with a high signal-to-noise ratio instead of lines,
which consequently excludes the regions in which azimuthal gradients are too small, below the observational noise.

When we do not assume $\vec n$ as known, we can determine the vector $\vec \Omega_\mathrm{LTW}$ by least-squares using three or more
distinct positions $\vec x_i$, $i=1,\ldots n$, for the 3D space density, and solve the equation system by linear optimization,
\begin{equation} \label{eq:LTWvector}
  \left(
    \begin{array}{c}
      {\vec x}_1 \wedge\partial_{\vec x}\rho({\vec x}_1)  \\
      \vdots \\
      {\vec x}_n \wedge\partial_{\vec x}\rho({\vec x}_n)  
    \end{array}
  \right) 
  \vec\Omega_\mathrm{LTW}  \approx  -         
  \left( 
    \begin{array}{c}
      \partial_{\vec x}\cdot \left( \rho({\vec x}_1)\,{{\vec u}}({\vec x}_1)\right)\\
      \vdots \\
      \partial_{\vec x}\cdot \left( \rho({\vec x}_n)\,{{\vec u}}({\vec x}_n)\right)
    \end{array} 
  \right) ,
\end{equation}
where 
\begin{equation}
\rho(\vec x,t) = \int \!d^3v \, f(\vec x,\vec v, t), 
\end{equation}
and
\begin{equation}
\rho(\vec x, t)\, \vec u(\vec x,t) = \int \!d^3v \, \vec v f(\vec x,\vec v, t).
\end{equation}
In principle, if these positions provide a nondegenerate set of linear equations, the full rotation vector
$\vec\Omega_\mathrm{LTW}$ can be retrieved.

Thus the LTW method consists in choosing a finite number of locations well in which the azimuthal density gradient
$\propto \left| {\vec x} \wedge\partial_{\vec x}\rho({\vec x}) \right|$ has a high signal-to-noise ratio.  In bars, the maximum
signal is to be sought in the off-axis regions, a butterfly shape, determined by preferentially accepting the data points where
$\left| {\vec x} \wedge\partial_{\vec x}\rho({\vec x}) \right|$ is above the noise.  Typically, locations near the principal axes of
a bar should be rejected: because the density gradient is radial, these regions add noise to the linear system.  Because this
involves measuring the gradient of $\rho$ and $\rho {\vec u,}$ it can be used in $N$-body bars where errors are under control.

\subsection{Regional method (RTW)}
However, if gradients are too noisy, we can proceed as in TW by integrating data over a volume. We do this here in 3D.  Instead of
the TW infinite line in the galaxy plane, we generalize by integrating over an arbitrary finite contiguous volume $V$ bounded by an
oriented surface $\vec S$ in one or several favorable high signal-to-noise regions.

The 3D local equation for the mass density equation $\rho(\vec x)$ in a velocity file $\bar{\vec v}(\vec x)$ is
\begin{equation} 
\left(\vec x \wedge \partial_{\vec x}\rho \right)\cdot {\vec  \Omega_\mathrm{RTW}} = 
  - \partial_{\vec x}\cdot \left( \rho \bar{\vec v}\right).  
\end{equation} 
Integrating over $V$, noting that $\nabla \rho \wedge \vec x = \nabla\wedge (\rho {\vec x})$, and using the divergence and curl
theorems of vector calculus%
\footnote{The lesser known curl theorem
$
\int_V d^3x\, \nabla \wedge \vec F({\vec x}) = \int_S d\vec S \wedge \vec F({\vec x})
$
for  a vector field $\vec F(\vec x)$ in a volume $V$ bounded by a surface $S$ is just a corollary of the divergence theorem
$
\int_V d^3x \, \nabla \cdot \vec F(\vec x) = \int_S d\vec S\cdot \vec F(\vec x),
$
by substituting $\vec F \to {\vec c} \wedge {\vec F}$ in the latter, where ${\vec c}$ is a constant vector.}, 
we obtain an integral over the surface $\vec S$ on both sides of the equation,
%
\begin{equation}\label{eq:RTWvector}
\left(\int_S \rho {\vec x}  \wedge  d{\vec S} \right) \cdot {\vec \Omega}_{RTW}  \approx  \int_S \rho\bar{\vec v}  \cdot d{\vec S} .
\end{equation}
These integrals amount to the total mass flux through surface $S$, one expressed with the supposed unknown uniform rotation of the
system%
\footnote{Since
  $\left(\rho {\vec x} \wedge d{\vec S}\right) \cdot {\vec \Omega}_\mathrm{RTW} = \rho \left({\vec \Omega}_\mathrm{RTW} \wedge {\vec
      x} \right)\cdot d{\vec S}$, and $\vec\Omega_\mathrm{RTW} \wedge \vec x $ is the tangential velocity.}, the other by the
measured mass flux parallel to $S$.  When  $S$ is chosen poorly, this flux may vanish in all cases, which is thus a degenerate
configuration to be avoided. For example, a sphere centered on the origin sees all $\vec x$ being parallel to the surface elements
$d\vec S$, so $(\rho {\vec x} ) \wedge d\vec S = 0$ for any $\rho$, Eq.\ (\ref{eq:RTWvector}) reduces to
$0 \cdot {\vec \Omega}_{RTW} =0$.

The RTW method can be considered to be a method that replaces the gradient of data evaluation at a point by directly integrating data over a
finite surface.  The optimal volume size is determined by the noise in the available data.  Since only the information present on
the surface of $V$ is used, there is no interest in taking overly large regions.

\subsection{Multiple regional method (MRTW)}
In principle, we can therefore combine the two previous approaches by a multiple RTW method (MRTW), for which several distinct
signal-rich regions inside volumes bounded by surfaces $S_1$, $S_2,\ldots S_n$ are used, allowing us to solve a series of Eq.\
(\ref{eq:RTWvector}) by linear optimization for $ {\vec \Omega_\mathrm{MRTW}}$,
\begin{equation} \label{eq:MRTWvector}
  \left(
    \begin{array}{c}
      \int_{S_1} \rho {\vec x}  \wedge  d{\vec S}_1 \\
      \vdots \\
      \int_{S_n} \rho {\vec x}  \wedge  d{\vec S}_n   
    \end{array}
  \right) 
  \vec\Omega_\mathrm{MRTW}  \approx           
  \left( 
    \begin{array}{c}
      \int_{S_1} \rho\bar{\vec v}  \cdot d{\vec S}_1 \\
      \vdots \\
      \int_{S_n} \rho\bar{\vec v}  \cdot d{\vec S}_n 
    \end{array} 
  \right) .
\end{equation}

For the Milky Way, using full 6D astrometric data from Gaia in volume-limited regions around the Sun where extinction is under
control, we might in principle attempt to evaluate the pattern speed of local spiral arms, and possibly more.



\section{Jacobi integral, or potential method}\label{sec:JACOBI}
Here we develop a method that does not require density gradients, but requires the individual kinematic data of test particles, or
stars.  This method was already briefly described in \cite{WPT18}.

In a gravitational mean potential $\Phi$ that rotates uniformly with angular frequency $\vec \Omega$, a test particle moving in this
potential always has the Jacobi integral $H_J$ as the global integral of motion \citep{Binney08},
\begin{equation}
H_J = {1\over 2} {\vec v}^2(t) + \Phi({\vec x}(t)) - {\vec \Omega} \cdot \left({\vec x}(t) \wedge {\vec v}(t)\right) 
    = e(t) -  {\vec \Omega} \cdot {\vec l}(t),
\end{equation}
where ${\vec x}(t)$ is the position of the particle, and ${\vec v}(t)$ the velocity in the inertial frame instantaneously
aligned with the rotating frame at time $t$. The quantity $e(t)= {1\over 2} {\vec v}^2(t) + \Phi({\vec x}(t))$ is the particle
specific energy in the inertial frame, and $\vec l(t)= \vec x(t) \wedge \vec v(t)$ is its specific angular momentum, also in the
inertial frame.

\subsection{Pattern speed knowing its direction}
When we know the direction of the rotation axis $\vec n$, $\vec \Omega = \Omega\,\vec n$, we can  in principle determine $\Omega$ from
the motion of a single particle because along an orbit
\begin{equation}
0 = \dot H_J(\vec x, \vec v)  = \dot e - \Omega \vec n \cdot \dot{\vec l} = \vec v\cdot \vec a +  \dot \Phi(\vec x) - \Omega \,\vec n 
\cdot (\vec x \wedge \vec a) , 
\end{equation}
that is, 
\begin{equation}\label{eq:omegaH}
\left(\vec x \wedge \vec a \right) \cdot \vec n\, \Omega = \vec v\cdot \vec a +  \dot \Phi(\vec x) . 
\end{equation}
In principle, provided the torque $\left(\vec x \wedge \vec a \right)$ is nonzero and not orthogonal to $\vec n$, a single particle
has the information about $\Omega$ in the values of its instantaneous position, velocity, acceleration and the time-derivative of
its potential along the trajectory, $\dot \Phi,$ at any time along its trajectory.  At some locations along a trajectory, the
tangential acceleration may become low (the torque term $\vec x \wedge \vec a$ in Eq.~(\ref{eq:omegaH}) almost vanishes). Then
$\Omega$ is ill defined. In nonaxisymmetric systems, these locations are exceptional.  Thus the best moments for determining
$\Omega$ along a trajectory occur when angular momentum varies the most, that is, when the torque is strong.

Alternatively, the above derivation leading to Eq.~(\ref{eq:omegaH}) can be obtained without invoking Jacobi's integral by assuming
that the potential $\Phi$ is an invariant function in some rotating frame, and using the equations developed in Sect.\
\ref{sec:ivfunc}.  Replacing $f$ by $\Phi$ in Eq.~(\ref{eq:omegainvfunc}), and noting that
$\dot \Phi = \partial_t \Phi + \vec v\cdot \partial_{\vec x}\Phi$, so $ \partial_t \Phi =\vec v\cdot\vec a +\dot \Phi $, we retrieve
Eq.~(\ref{eq:omegaH}), with $\vec n \Omega$ replaced by $\vec \Omega$.
 
\subsection{Pattern speed vector}
Using three or more independent particles, we can form the linear system
\begin{equation}
    \underbrace{
    \left(
    \begin{array}{c}
      \vec x_1 \wedge \vec a_1 \\
      \vdots\\
      \vec x_n \wedge \vec a_n
    \end{array}
  \right)
  }_{n\times 3 \ \textrm{\scriptsize matrix}}
  \vec\Omega
  \approx 
  \underbrace{
  \left(
    \begin{array}{c}
      \vec v_1\cdot \vec a_1 + \dot \Phi(\vec x_1) \\
      \vdots\\
      \vec v_n\cdot \vec a_n + \dot \Phi(\vec x_n) 
    \end{array}
    \right)    
}_{n\times 1 \ \textrm{\scriptsize vector}}
,
\end{equation}
which can be solved for $\vec \Omega$ by linear optimization. Furthermore, as in Eq.\ (\ref{eq:omegaVA}), we can extend the search
for the motion of the system rotation center at speed $\vec V$,
\begin{equation}
    \underbrace{
    \left(
    \begin{array}{cc}
      \vec x_1 \wedge \vec a_1 & \vec a_1 \\
      \vdots & \vdots  \\
      \vec x_n \wedge \vec a_n & \vec a_n\\
    \end{array}
  \right)
  }_{n\times 6 \ \textrm{\scriptsize matrix}}
  \underbrace{
    \left(
  \begin{array}{c}
  \vec\Omega \\
  \vec V \\
  \end{array}
  \right)
  }_{6\times 1 \ \textrm{\scriptsize matrix}}
  \approx 
  \underbrace{
  \left(
    \begin{array}{c}
      \vec v_1\cdot \vec a_1 + \dot \Phi(\vec x_1) \\
      \vdots\\
      \vec v_n\cdot \vec a_n + \dot \Phi(\vec x_n) 
    \end{array}
    \right)    
}_{n\times 1 \ \textrm{\scriptsize vector}}
.
\end{equation}

The particularity of this method is that $\dot\Phi$ must be determined.  In usual $N$-body simulation codes, all the other values
except for $\dot \Phi$ are calculated at each time step. With a small increase in computation cost, the force algorithm can also
calculate $\Phi$, so that $\dot \Phi(t)$ can be approximated to second order in $\Delta t$ by finite difference in time and two
additional potential calculations,
\begin{equation}
  \dot \Phi(\vec x(t))  =   {\Phi(\vec x(t+\Delta t))-\Phi(\vec x(t-\Delta t)) \over 2\Delta t} + {\cal O}(\Delta t^2) 
   \approx {\Phi(\vec x(t) +\vec v(t) \Delta t)-\Phi(\vec x(t) - \vec v(t) \Delta t ) \over 2\Delta t}   ,
\end{equation}
where $\Delta t$ is a positive time step.  In a pure $N$-body system, and similarly, for other potential solvers, the time
derivative of the potential can be calculated directly with little additional expense simultaneously with the potential and forces
by
\begin{equation}
  \dot \Phi(\vec x_j(t)) = {d\over dt}\left( -G \sum_{i\neq j} {m_i \over |{\vec x}_i(t)-{\vec x}_j(t)|} \right)
 = G \sum_{i\neq j}  m_i {({\vec x}_i(t)-{\vec x_j}(t))\cdot({\vec v}_i(t)-{\vec v}_j(t)) \over  |{\vec x}_i(t)-{\vec x}_j(t)|^3} .
\end{equation}
In an $N$-body system, however, the potential is clearly never strictly time independent. It differs from a mean potential because
nearby particles eventually contribute to strong fluctuations of the potential over the background of slowly varying contributions
from particles that are far away.  The distribution of the found $\vec \Omega$ obtained from a set of selected particles can help to
quantify the force fluctuations from the nearest particles with respect to the smoother mean field force: if the distribution of
found $\vec \Omega$ is broad, it means that there is no single frame with a uniformly rotating potential, or that the fast local
fluctuating forces dominate the more constant force from the mean rotating potential. In a smooth mass distribution, the
contribution to force fluctuations from nearby particles becomes particularly important near the potential minimum (near the
center), or everywhere in extended nearly homogeneous distributions like in cosmology.


\section{Second-order moment  methods}\label{sec:MOMENT}
\subsection{Moment of inertia tensor}

In mechanics, the polar moment of inertia tensor ${\vec I}$ is a simple $3\times 3$ tensor characterizing the second moments of the
mass distribution with respect to the origin. It is set at the location at which the center of mass and center of rotation coincide.
It can be seen as the covariance matrix weighted by the mass of the particle positions.  For an $N$ particle distribution, it is
defined as
\begin{equation}
  {\vec I} \equiv   {\vec X}^T {\vec M} {\vec X} =           
   \left( 
      \begin{array}{ccc}
        \sum_i m_i x_i^2   &  \sum_i m_i x_i y_i  &  \sum_i m_i x_i z_i \\
        \sum_i m_i x_i y_i &  \sum_i m_i y_i^2    &  \sum_i m_i y_i z_i \\
        \sum_i m_i x_i z_i &  \sum_i m_i y_i z_i  &  \sum_i m_i z_i^2 
      \end{array} 
   \right) 
   ,
\end{equation}
where $ {\vec X}$ is the $N\times 3$ matrix composed of the stacked particles in Cartesian coordinates ${\vec x}_i$,
\begin{equation}
  {\vec X} \equiv     \left( 
      \begin{array}{ccc}
        x_1   &  y_1 & z_1 \\
        \vdots & \vdots & \vdots \\
        x_N  &  y_N & z_N 
      \end{array} 
   \right) ,
\end{equation}
and ${\vec M}$ is the $N\times N$ mass diagonal matrix, with its diagonal filled with the particle masses $m_i$,
\begin{equation}
  {\vec M} \equiv   
   \left( 
      \begin{array}{cccc}
        m_1   
                    & \dots & 0 \\
        \vdots
                    & \ddots & \vdots \\
        0 
                    & \dots & m_N 
      \end{array} 
   \right) .
\end{equation}
When only the average particle distribution is rigid except for a uniform rotation about an axis, we can still use the axial moment
of inertia tensor in use in solid body mechanics \citep[e.g.,][]{Goldstein71} even though each particle moves nonrigidly with
respect to each other to relate the rotation of the particle distribution to its constant total angular momentum vector
$\vec L = \sum_i m_i \vec x_i \wedge \vec v_i $ and its axial moment of inertia tensor, which is given by
\begin{equation}
  \mathbb{I} \equiv   \mathop{\mathrm{Tr}}({\vec I}){\vec 1}  -  {\vec I}   
              = 
  \left( 
      \begin{array}{ccc}
        \sum_i m_i (y_i^2+z_i^2)   &  -\sum_i m_i x_i y_i  &  -\sum_i m_i x_i z_i \\
        -\sum_i m_i x_i y_i &  \sum_i m_i (z_i^2+x_i^2)    &  -\sum_i m_i y_i z_i \\
        -\sum_i m_i x_i z_i &  -\sum_i m_i y_i z_i  &  \sum_i m_i( x_i^2+y_i^2) 
      \end{array} 
   \right) 
. 
\end{equation}
Then the angular speed $\vec \Omega_\textrm{s}$ of the solid body is related to $\vec L$ by
\begin{equation} \label{equ:solidrot}
  \vec L = \mathbb{I} {\vec \Omega_\textrm{s}}. 
\end{equation}
As the system rotates, $\vec I$ and therefore $\mathbb{I}$ vary: the non-diagonal terms, for example, vanish when the Cartesian
coordinates are aligned with the principal axes of the mass distributions.  $\vec \Omega_\textrm{s}$ must still be constant,
however, because $\vec L$ is conserved and its value should not change under a rotation of the coordinates about $\vec L$.  This
means that knowing the particle distribution and its total angular momentum allows us to find a unique ${\vec \Omega_\textrm{s}}$
that describes the instantaneous rotation rate as if the particles were moving as a solid.  When the particle distribution does not
rotate rigidly, then ${\vec \Omega_\textrm{s}}$ can clearly change, still respecting Eq.~(\ref{equ:solidrot}).

The only technical problem is to find a well-behaved invert of $\mathbb{I}$ for all possible configurations. \cite{pfenniger2019}
gave an explicit procedure for finding the inverse $\mathbb{I}^{-1}$, or more generally, the pseudo-inverse $\mathbb{I}^{\dagger}$
in degenerate cases,
\begin{equation}
  \vec\Omega_\textrm{s} = \mathbb{I}^\dagger {\vec L}. 
\end{equation}
For a body with internal streaming, however, such as a bar in a galaxy, $\vec \Omega_\textrm{s}$ is not necessarily the pattern
speed because its value does not depend on any feature of the mass distribution and also works with axisymmetric systems rotating
about the symmetry axis.

The polar moment of inertia tensor ${\vec I}$ described above is a simple $3\times 3$ tensor characterizing the second moments of
the mass distribution with respect to the origin, set at the center of mass. If the system globally keeps its shape over time while
rotating uniformly, the eigenvalues of ${\vec I}$ remain constant, and its rotating eigenvectors may be used to determine the
pattern speed.

\subsubsection{Pattern speed of the 3D system vector}\label{sec:MOMENT_I}
In principle, the real eigenvalues and eigenvectors can be calculated analytically for a 3x3 symmetric matrix, but the expressions
are huge because the eigenvalues are solutions of a cubic polynomial.  It is more practical to use well-established numerical
methods to determine them \citep{lapack99,TrefethenBau1997}.

For a symmetric positive definite matrix $A$, the eigenvalue problem, which consists of finding the orthogonal eigenvector matrix
$V$ and the diagonal eigenvalue matrix $\Lambda$ for which $A = V\Lambda V^T$, is identical to the singular value decomposition
(SVD) problem for a general (nonsquare) matrix: $A = U S V^T$, where $S$ is the diagonal matrix of (real) singular values and $U$
and $V$ are two orthogonal matrices.  For positive definite symmetric matrices $U=V$, the eigenvalue problem is thus equivalent to
the SVD problem.  In contrast to standard SVD algorithms, however, common eigenvalue--eigenvector algorithms do not sort eigenvalues
and eigenvectors in a standard way because they may be complex or real negative.  It is therefore more convenient to use the SVD
algorithm.  For small matrices, the difference in computational cost is negligible.

After decomposing ${\vec I}$ with SVD: ${\vec I} = USU^T$, we obtain a real orthogonal rotation matrix $U$ made of the three
orthogonal unit vectors $\vec n_i$, where $i=1, 2, 3$, $U = [\vec n_1, \vec n_2, \vec n_3]$, which gives the direction of the
eigenvectors, and a diagonal real matrix $S$ with singular values $s_1 \geq s_2 \geq s_3\geq 0$ in decreasing order, giving the
second-order moment amplitudes in the respective directions.  If the second-order moments are invariant, $S$ does not depend on
time, but $\dot U$, the time derivative of $U$, is a matrix giving the speed of the eigenvectors.

To evaluate $\dot U$, we first need  $\dot {\vec I}(t)$, which is obtained using the particle positions $\vec X$ and velocities
$\vec V$ (assuming mass conservation, $\dot {\vec M} = 0$),
\begin{equation}
 \dot {\vec I} = \dot {\vec X}^T {\vec M} {\vec X} + {\vec X}^T {\vec M} \dot {\vec X} 
 ~ = ~ {\vec V}^T {\vec M} {\vec X} + {\vec X}^T {\vec M} {\vec V}  
  =           
   \left( 
      \begin{array}{ccc}
        2 \sum_i m_i x_i v_{x_i}   &  \sum_i m_i (x_i v_{y_i} + y_i v_{x_i})  &  \sum_i m_i (x_i v_{z_i}  + z_i v_{x_i} ) \\
         \sum_i m_i (x_i v_{y_i} + y_i v_{x_i}) &  2\sum_i m_i y_i v_{y_i}    &  \sum_i m_i (y_i v_{z_i}  + z_i v_{y_i} ) \\
        \sum_i m_i (x_i v_{z_i}  + z_i v_{x_i} ) & \sum_i m_i (y_i v_{z_i}  + z_i v_{y_i} )  & 2 \sum_i m_i z_i v_{z_i} 
      \end{array} 
   \right) . 
\end{equation}

A safe, accurate method for calculating the derivative of the SVD, knowing $\dot{\vec I}$ is described in Appendix A.  To do this,
knowing the SVD decomposition of $\dot{\vec I}$, the symmetric matrix $H = U^T\dot{\vec I} U $ is first calculated.  Its diagonal is
the diagonal of $\dot S$, while its nondiagonal elements allow us to build an auxiliary antisymmetric matrix $\tilde U$.  The time
derivative of $U$ is then found by $\dot U = U \tilde U$.  The singular value derivative matrix $\dot S$ allows us to check the
radial rigidity of the rotating pattern, which is small by hypothesis. Along the respective eigenvector $i,$ each $\dot s_i$
provides a check of rigidity, and $s_i/\dot s_i$ a timescale for substantial radial change.

Then, using the unit vectors $\vec n_i$ and $\dot{\vec n}_i$, $i=1,\ldots 3$, of $U = \left( \vec n_1, \vec n_2, \vec n_3 \right)$
and $\dot U = (\dot{\vec n}_1, \dot{\vec n}_2, \dot{\vec n}_3),$ respectively, corresponding to the singular values $s_i$, we derive
the instantaneous rotation frequency vector of each of the principal axes along $\vec n_i$,
\begin{equation}
   \vec \Omega_{I_i}  = \vec{n}_i(t) \wedge \dot{\vec n_i}(t), \quad i=1,\ldots 3.
\end{equation}
If the particle ensemble rotates as a rigid figure, and if the axis of rotation is not aligned with any $\dot{\vec n}_i$, the three
$\vec \Omega_{I_i} $ should be almost identical within the numerical errors.  However, figures frequently rotate close to around one
of the principal axes, in which case the corresponding $|\dot{\vec n_j}|$, thus $|\vec \Omega_{I_j}|$, is small or zero.  In the
frequent case that the figure rotates around the shortest axis, $\vec \Omega_{I_1}$ and $\vec \Omega_{I_2}$ are almost identical and
provide the sought pattern speed.  Determining the instantaneous vector $\vec \Omega_I$ of distinct subsets of particles allows us
to quantify as a function of time by how much the rotation axes of these subsets may tilt with respect to each other.

\subsubsection{Pattern speed of the 2D system }
When the direction of the rotation axis is known and constant, for instance, along the $z$-axis, we can evaluate $\Omega$ explicitly
avoiding SVDs.  Noting
\begin{eqnarray}\label{equ:moments}
  &&I_{xx} = \sum_i m_i {x}_i^2, \quad  I_{yy} = \sum_i m_i {y}_i^2, \quad  I_{xy} = \sum_i m_i {x}_i{y}_i, \quad
    I_{\pm} = {1\over 2} \left(I_{xx} \pm I_{yy}\right) = {1\over 2} \sum_i m_i \left({x}_i^2 \pm {y}_i^2\right), \nonumber \\
  &&\dot I_{xy} = \sum_i m_i \left({x}_i {v_y}_i+{y}_i{v_x}_i\right), \quad
  \dot I_{-} = \sum_i m_i \left( {x}_i {v_x}_i - {y}_i {v_y}_i\right),\quad
   P = \sqrt{I_{-}^2+I_{xy}^2},
\end{eqnarray}
we calculate the eigenvalues $\lambda_\pm$ by solving the characteristic polynomial of $\vec I$, 
\begin{equation}
0 = \left| \vec I -\lambda \vec 1\right| = \lambda^2 - 2I_+ \lambda + I_+^2- P^2, 
\end{equation}
which gives two solutions for $\lambda$: $\lambda_\pm = I_+ \pm P$.  
The eigenvalue ratio is a measure of the moment-of- inertia ellipsoid axis-ratio squared, thus the ellipticity reads
\begin{equation}
\epsilon = \sqrt{\lambda_- \over \lambda_+} = \sqrt{I_+-P \over I_++P}. 
\end{equation}

The eigenvectors $\vec \xi_\pm$ only depend on the ratio $\zeta \equiv I_{xy}/I_{-}$ ,  
\begin{equation}
 \vec \xi_\pm = \left\{ \pm \zeta, \sqrt{  1 + \zeta^2}\mp 1 \right\} . 
\end{equation}
The two vectors can be verified to be orthogonal, that is, $\xi_+\cdot \xi_- = 0$.  Using identity 1.626.3 in
\cite{GradshteynRyzhik2007} \footnote{$2\arctan(x) = \arctan\left(2x\over 1-x^2\right) + \Heaviside(x^2-1)\,\sgn(x)\,\pi$, where
  $\Heaviside(x)$ is Heaviside's step function.}, the angle $\phi$ at time $t$ between one of the eigenvectors and the $x$-axis in
the $x-y$ plane is
\begin{equation}
  \phi (t) =  { 1 \over 2}\arctan\left(\zeta\right) + \mathrm{constant}. 
\end{equation}
The constant is $0$, or $\pm \pi/2$, which is irrelevant for our aim because it disappears when we differentiate $\phi$ with respect
to time. We find
\begin{equation}
  \label{eq:omegaz}
  \Omega_{Iz} \equiv \dot \phi (t) = {1\over 2} {\dot \zeta \over 1+\zeta^2} = {1\over 2} { I_{-}\dot I_{xy} - \dot I_{-}I_{xy} \over I_{-}^2 + I_{xy}^2}  .
\end{equation}
The instantaneous angular speed of the moment of inertia at any time can therefore be calculated directly with the position and
position-velocity moments given in Eq.\ (\ref{equ:moments}).

When the particle distribution is close to axisymmetric, for instance, $I_{xy} \approx 0 $ and $ I_{-} \approx 0$, $\Omega_z $ is
undetermined, as it should.  This also occurs when the distribution, while not axisymmetric, has strictly no bisymmetric component,
however, such as a pure $\text{four}-$arm spiral pattern.  This method was briefly described in \cite{WPT18}.

\subsubsection{Pattern acceleration of the 2D system}
The time derivative of $\Omega_{Iz}$ can be calculated similarly, knowing the particle accelerations, which are calculated in
$N$-body simulations in any case.  This can be useful for estimating the timescale over which the pattern speed can be approximated
as constant.  Differentiating Eq.~(\ref{eq:omegaz}), we obtain the relative pattern acceleration after simplifications,
\begin{equation}
  \label{eq:reldotomegaz}
  {\dot \Omega_{Iz} \over \Omega_{Iz}} = 
       {
         I_{-} \ddot I_{xy} - I_{xy} \ddot I_{-}
         \over
         I_{-} \dot I_{xy}  - I_{xy} \dot I_{-}
       }
       - 2\, 
       {I_{-} \dot I_{-} + I_{xy} \dot I_{xy}
         \over
         I_{-}^2 + I_{xy}^2 
       }, 
\end{equation}
where
\begin{eqnarray}
  \ddot I_{xy} ~=~ \sum_i m_i \left( 2{v_x}_i{v_y}_i + x_i {a_y}_i + y_i {a_x}_i\right), \quad
  \ddot I_{-} ~=~ \sum_i m_i \left({v_x}_i^2 - {v_y}_i^2 + x_i {a_x}_i - y_i {a_y}_i\right).
\end{eqnarray}

Dimensionless estimators of the pattern speed invariance can be obtained by dividing Eq.\ (\ref{eq:reldotomegaz}) by a reference
frequency, such as the solid-body speed $\Omega_\textrm{s}$, or the pattern speed $\Omega_{Iz}$ itself.  When the dimensionless
estimator is of order unity or more, the constant pattern speed assumption is no longer valid.

\subsection{Kinetic tensor rotation}
The velocity distribution can rotate in velocity space with an angular speed that is distinct from the angular speed in position
space.  Exactly the same discussion as in Sect.\,\ref{sec:MOMENT_I} for the inertia tensor method can be carried out for the
second-order moment tensor of velocities, the kinetic tensor (where the velocity origin is set at the velocity center of mass),
\begin{eqnarray}
  {\vec K} &\equiv &  {\vec V}^T {\vec M} {\vec V} =           
   \left( 
      \begin{array}{ccc}
        \sum_i m_i {v_x}_i^2   &  \sum_i m_i {v_x}_i {v_y}_i  &  \sum_i m_i {v_x}_i {v_z}_i \\
        \sum_i m_i {v_x}_i {v_y}_i &  \sum_i m_i {v_y}_i^2    &  \sum_i m_i {v_y}_i {v_z}_i \\
        \sum_i m_i {v_x}_i {v_z}_i &  \sum_i m_i {v_y}_i {v_z}_i  &  \sum_i m_i {v_z}_i^2 
      \end{array} 
   \right) .
\end{eqnarray}
All the calculations for finding the 3D pattern speed with SVD decomposition can be made to obtain the pattern speed of the velocity
ellipsoid by replacing positions by velocities, and velocities by accelerations, as in Sect.\ref{sec:MOMENT_I}. We therefore do not
elaborate further.

Likewise, for the 2D case, using these notations, 
\begin{eqnarray}
 && K_{xx} ~=~ \sum_i m_i {v_x}_i^2, \quad  K_{yy} = \sum_i m_i {v_y}_i^2, \quad  K_{xy} = \sum_i m_i {v_x}{v_y}_i, \quad
  \dot K_{xy} ~=~ \sum_i m_i \left({v_x}_i {a_y}_i+{v_y}_i{a_x}_i\right), \nonumber \\
 &&  E_{xy} ~ = ~ {1\over 2}\left(K_{xx} - K_{yy}\right) \ = \  {1\over 2} \sum_i m_i \left({v_x}_i^2-{v_y}_i^2\right), \quad
  \dot E_{xy} ~=~ \sum_i m_i \left( {v_x}_i {a_x}_i - {v_y}_i {a_y}_i\right),
\end{eqnarray}
we obtain the pattern speed of the velocity ellipsoid,
\begin{equation}
  \label{eq:omegazK}
  \Omega_{Kz} = {1\over 2} { E_{xy}\dot K_{xy} - \dot E_{xy}K_{xy} \over E_{xy}^2 + K_{xy}^2}.
\end{equation}

To evaluate the pattern acceleration, we would need to know the particle jerk ($\dot{\vec a}$). 

\section{Fourier method}\label{sec:FOURIER}

A 2D mass distribution can have a simple angular Fourier decomposition where a low-order mode may capture the nonaxisymmetric shape
of the distribution.  This is commonly used to measure the strength, phase, and length of a bar, for instance, where the dominant
$m=2$ mode is often exclusively considered, while the $m=4$, 6 and 8 modes are still non-negligible.  The $m=2$ and higher-order
modes are thought to rotate at the same angular velocity as the mass distribution.  In this case, we can attempt to determine the
pattern speed by measuring just the mode phase velocity.
  
Below we describe the Fourier method introduced in \cite{WPT18} more extensively.  An inverse Fourier decomposition in the plane
$x-y$ of the mass-weighted azimuthal positions of a distribution of particles sampled in a ring reads
\begin{equation}
  F_m = \sum_j m_j \exp(\mathrm{i}  m \theta_j) =\sum_j m_j \left(\cos(m \theta_j)+\mathrm{i}\sin(m \theta_j)\right) , 
\end{equation}
where the azimuth angle $\theta_j$ of particle $j$ depends on the Cartesian coordinates $x_j$, and $y_j$,
$\theta_j = \arctan(y_j,x_j)$, and $m$ is the Fourier mode.  The summation runs over the number of particles in the ring. The phase
of the Fourier transform is
\begin{equation}
  \phi_m = \arctan(\Im(F_m),\Re(F_m)),
\end{equation}
where
\begin{eqnarray}
  \Re(F_m) = \sum_j  m_j \cos( m \theta_j), &&
  \Im(F_m) = \sum_j m_j \sin( m \theta_j).
\end{eqnarray}

\subsection{2D Fourier mode velocity}
Because the mode phase goes from 0 to $2\pi$ when the particle positions rotate in space by $2\pi/m$, the mode velocity in space is
the phase velocity divided by $m$,
\begin{equation}\label{eq:omegam}
  \Omega_{m} \equiv 
    {\dot{\phi}_m\over m} = {1\over m} {\dot\Im(F_m) \Re(F_m) - \dot\Re(F_m) \Im(F_m) \over\Re(F_m)^2 + \Im(F_m)^2} ,  
\end{equation}
where 
\begin{eqnarray}
{\dot\Re(F_m)\over m} = -\!\sum_j m_j \sin( m \theta_j) \, \dot \theta_j, && 
{\dot\Im(F_m)\over m} = \sum_j m_j \!\cos( m \theta_j) \, \dot \theta_j,
\end{eqnarray}
and
\begin{equation}
  \dot \theta_j = {x_j{\dot y}_j - y_j {\dot x}_j \over x_j^2+y_j^2} =  {x_j{v_y}_j - y_j {v_x}_j \over x_j^2+y_j^2}. 
\end{equation}
Explicitly, defining the following trigonometric sums,
\begin{eqnarray}\label{eq:CSsums}
 && C    \equiv \sum_j m_j \cos(m\theta_j),             \quad S    \equiv \sum_j m_j \sin(m\theta_j), \quad
 C_1  \equiv \sum_j m_j \cos(m\theta_j)\, \dot\theta_j, \quad S_1  \equiv \sum_j m_j \sin(m\theta_j)\, \dot\theta_j,
\end{eqnarray}
we obtain the compact expression,
\begin{equation}\label{eq:Omegam}
  \Omega_{m} = {C C_1 + S  S_1 \over C^2 + S^2 }.
\end{equation}

When several modes superpose with different mode velocities, the mass distribution is time-dependent and no single pattern speed
characterizes the distribution well.  Determining how well the large amplitude $(C^2+S^2)^{1/2}$ mode velocities coincide is
therefore a way to verify the time-independence of the pattern.

\subsection{Fourier mode acceleration}
Another way to confirm time dependence is to calculate the instantaneous mode acceleration.  Differentiating Eq.~(\ref{eq:omegam}),
and defining
\begin{eqnarray}
  && C_2 \equiv \sum_j m_j \left[ \cos( m \theta_j)\,\ddot\theta_j - m \sin( m \theta_j)\,\dot\theta_j^2\right],\quad
     S_2 \equiv \sum_j m_j \left[ \sin( m \theta_j)\,\ddot\theta_j + m \cos( m \theta_j)\,\dot\theta_j^2\right],
\end{eqnarray}
with 
\begin{equation}
  \ddot \theta_j =  {\left(x_j{a_y}_j - y_j {a_x}_j\right) - 2 \dot \theta_j \left(x_j{v_x}_j + y_j {v_y}_j\right) \over x_j^2+y_j^2},   
\end{equation}
we obtain the relative mode acceleration,
\begin{equation}
  {\dot \Omega_m \over \Omega_m} = {C C_2 + S  S_2 \over  C C_1 + S  S_1} + 2m {C S_1 - S C_1 \over C^2 + S^2 }.
\end{equation}
Thus we can also calculate the instantaneous variation of
the mode phase velocity from the particle coordinates, velocities, and accelerations at any time.

\subsection{3D extension for the $m=2$ mode}
Extending a Fourier-like decomposition for 3D distributions in order to determine a given mode speed vector is not straightforward.
For example, choosing a decomposition in spherical harmonics or a double Fourier expansion over the sphere predetermines the
positions of the poles and therefore breaks isotropy.

For the $m=2$ mode in 2D, however, the Fourier method is equivalent to the moment of inertia method, as shown by substitution, where
the positions $\vec x_i$ are replaced by unit vectors $\vec n_i = \vec x_i/\|\vec x_i\| $.  We showed above that the moment of
inertia method works for 3D naturally using a SVD decomposition and differentiation. We can therefore use this method to replace the
position vectors by unit vectors weighted by the mass.

The angular moment of inertia tensor ${\vec I_n}$ reads
\begin{eqnarray}
  {\vec I_n} ~\equiv~  {\vec N}^T {\vec M} {\vec N} ~=~           
   \left( 
      \begin{array}{ccc}
        \sum_i m_i n_{x_i}^2  &  \sum_i m_i n_{x_i} n_{y_i}  &  \sum_i m_i n_{x_i} n_{z_i} \\
        \sum_i m_i n_{x_i} n_{y_i} &  \sum_i m_i n_{y_i}^2    &  \sum_i m_i n_{y_i} n_{z_i} \\
        \sum_i m_i n_{x_i} n_{z_i} &  \sum_i m_i n_{y_i} n_{z_i}  &  \sum_i m_i n_{z_i}^2 
      \end{array} 
    \right) 
   ,
\end{eqnarray}
where $ {\vec N}$ is the $N\times 3$ matrix composed of the particles unit vector coordinates
${\vec n}_i = {\vec x}_i / \| {\vec x}_i \| $,
\begin{eqnarray}
  {\vec N} &\equiv &    \left( 
      \begin{array}{ccc}
        n_{x_1}   &  n_{y_1} & n_{z_1} \\
        \vdots & \vdots & \vdots \\
        n_{x_N}  &  n_{y_N} & n_{z_N} 
      \end{array} 
   \right) .
\end{eqnarray}
To determine $\dot{\vec N}$, we first calculate the time derivatives of a unit vector $\vec n$,
\begin{equation}
  {\vec u} ~ \equiv ~\dot{\vec n} ~=~ {\dot{\vec x}\over \|\vec x\|} - {\vec x \cdot \dot{\vec x}\over \|\vec x\|^2}
               ~=~  {\vec v\over \|\vec x\|} - {\vec n \cdot \vec v \over \|\vec x\| }\vec n,
\end{equation}
which is just the angular speed of the unit vector $\vec n$.  Therefore,
\begin{eqnarray}
  {\vec U} \equiv \dot{\vec N } \equiv \left( 
      \begin{array}{ccc}
        u_{x_1}   &  u_{y_1} & u_{z_1} \\
        \vdots & \vdots & \vdots \\
        u_{x_N}  &  u_{y_N} & u_{z_N} 
      \end{array} 
   \right) .
\end{eqnarray}
Thus
\begin{eqnarray}
  && \dot {\vec I_n} \equiv  {\vec U}^T {\vec M} {\vec N} + {\vec N}^T {\vec M} {\vec U} =     
  \sum_i\left( 
      \begin{array}{ccc}
         2 m_i n_{x_i}u_{x_i}   &    m_i (n_{x_i}u_{y_i} +  n_{y_i}u_{x_i})   &    m_i (n_{x_i}u_{z_i}  +  n_{z_i}u_{x_i})   \\
         m_i (n_{x_i}u_{y_i}  +  n_{y_i}u_{x_i})   &   2 m_in_{y_i}u_{y_i}    &    m_i (n_{y_i}u_{z_i}  +  n_{z_i}u_{y_i})   \\
         m_i (n_{x_i}u_{z_i}  +  n_{z_i}u_{x_i})   &    m_i (n_{y_i}u_{z_i}  +  n_{z_i}u_{y_i})   &    2m_i n_{z_i}u_{z_i}  
      \end{array} 
    \right) 
   .
\end{eqnarray}
Knowing $\dot {\vec I_n} $ and the SVD decomposition of ${\vec I_n}$ allows us to proceed exactly as in Sect.~\ref{sec:MOMENT}, therefore, we
do dot elaborate further.


\section{Conclusions}\label{sec:CONCLUSIONS}
We have presented five different formal methods for determining the instantaneous pattern speed of sets or subsets of particles, or
of stars, under the assumption that at the broad scale, these sets or subsets possess gradients that rigidly rotate around some
axis. This is only a first approximation of the more complex reality in simulations and in galaxies.

The methods include checks that the pattern acceleration is low, or that different modes in a Fourier decomposition rotate at the
same speed.  The local methods are useful for probing the spatial variations of instantaneous local pattern speeds in real
situations, allowing us to estimate the intrinsic errors made in assuming a constant global pattern speed.  Some methods require
knowing, or estimating, the particle accelerations, which for the present time are difficult to measure in the Milky Way, but should
become increasingly more available in the future. Accelerations in the Milky Way can currently best be inferred from mass modeling
and by solving Poisson's equation.

We have already used and successfully tested some of these methods in published \citep{WPT16,WPT18} or unpublished works.  In a
sequel paper, we plan to present tests, checks, and code of the most useful methods we presented here.

\begin{acknowledgements}
  DP thanks both the Inter-University Centre for Astronomy and Astrophysics, Pune, India, and the Academia Sinica Institute of
  Astronomy and Astrophysics, Taipei, Taiwan, for hospitality for a number of stays where this work was extended. KS thanks Geneva
  observatory for the kind hospitality during 2013 and subsequent years when the initial phase of this work started.  YTW thanks
  Geneva observatory for hospitality for a number of visits in 2015 and subsequent years when this work was extended.
\end{acknowledgements}

\section*{Notes added in proof}
\begin{enumerate}  
\item An older web publication, besides the ones cited in Appendix A, by Mike Giles\footnote{\textit{An extended collection of
      matrix derivative results for forward and reverse mode algorithmic differentiation}, Mike Giles, (2008),
    \texttt{https://people.maths.ox.ac.uk/gilesm/files/NA-08-01.pdf}}, has been found which describes, among others, the derivatives
  of eigenvalue and singular value decompositions, including computer code.  In particular the eigenvalue decomposition derivative
  is an equivalent but slightly more efficient method than the SVD method presented here.

\item The initial phases of our work, including the original moment and Fourier methods, was openly
  presented many times in seminars and workshops, such as the 2013, 2014 and 2016 Gaia Challenge
  Workshops\footnote{\texttt{http://astrowiki.ph.surrey.ac.uk/dokuwiki/doku.php?id=tests:discs}}. The methods have been developed over the years, and some of the methods briefly described and applied in \cite{WPT16,WPT18}.  
  While submitting this article \cite{Dehnen_etal2023} published a paper using very similar Fourier and moment methods
  to Sect.\ \ref{sec:MOMENT}-\ref{sec:FOURIER}.  
  They add a smoothing weight function to their derivations, while keeping our $C$ and $S$ notations used in \cite{WPT18} and in Eq.\ (\ref{eq:CSsums}).
  This is not a coincidence, because we publicly disclosed our single snapshot Fourier and moment methods years before.  
  
  Their weight function correction, using the additional mass conservation assumption, may be a useful option to add to the moment and Fourier methods in Sects.\ \ref{sec:MOMENT}--\ref{sec:FOURIER} whenever sharp windowing of particles is used.  However, in a particular N-body simulation we have found that the radial flux correction for a ring containing less than about $10^5$ particles was an illusory improvement because the noise introduced by the mass flux correction terms was comparable to the intended bias correction. 
  This calls for more testing of the respective approaches. 
\end{enumerate}

\bibliographystyle{aa}

\bibliography{biblio}


\section*{Appendix A: SVD differentiation}

For any matrix $A$, here illustrated for a $3\times3$ matrix, the singular value decomposition
\begin{equation}\label{eq:A}
  A = U S V^T,
\end{equation}
consists of two orthogonal matrices $U$ and $V$, and a diagonal real matrix $S$ where the non-negative singular values $s_1$,
$s_2, \ldots$ are sorted in decreasing order $s_1\geq s_2 \geq\ldots$ along the diagonal,
\begin{equation}\label{eq:S}
  S =
  \left(
    \begin{array}{ccc}
      s_{1} & 0 & 0 \\
      0 & s_{2} & 0 \\
      0 & 0 & s_{3}
    \end{array}
  \right).
\end{equation}
The goal here is to explicitly describe the behavior of the SVD when the matrix $A$ is perturbed to $A + dA$.  We have found very
little useful literature on this topic.  The derivation below is based on web publications by Ladislav
Kavan\footnote{\textit{PBA2014: SVD differentiation}, Ladislav Kavan, 2014, \tt https://www.youtube.com/watch?v=1cCwKsgsFX8} and
James Townsend\footnote{\textit{Differentiating the Singular Value Decomposition}, James Townsend, 2016, \tt
  https://j-towns.github.io/papers/svd-derivative.pdf}.

For symmetric (square) matrices, the singular-value problem is the same as the eigenvalue problem $AV = VS$ because then $U=V$.  We
therefore show the case for general $3\times 3$ matrices below.  Townsend's article contains the matrix expressions for the fully
general rectangular case of any size.

Differentiating Eq.\ (\ref{eq:A}),
\begin{equation}
  dA = dU\, S\, V^T + U\, dS\, V^T + U\, S\, dV^T,
\end{equation}
and multiplying left and right by $U^T$ and $V$, we obtain
\begin{equation}\label{eq:UTdAV}
  U^T dA\,V =  dS + (U^T dU) S + S(V^TdV)^T.
\end{equation}
When $A$, $U$, $V$, $S$, and $dA$ are known, the task is now to explicit the differential part of the SVD decomposition, namely
$dS$, $dU$, and $dV$.  The orthogonality of $U$ and $V$ ensures that $U^T U = 1$, and $V^T V = 1$. Differentiating $U^T U$,
\begin{equation} \label{eq:dU}
  U^T dU + dU^T U = U^T dU + (U^T dU)^T = 0, 
\end{equation}
we see that the matrix $U^TdU$ is antisymmetric. The same holds for $V^TdV$.  Naming $\tilde U = U^TdU$ and $\tilde V = V^TdV$, and
writing element-wise,
\begin{equation}
  \tilde U =
  \left(
    \begin{array}{ccc}
                  0 &  \tilde u_{12} & -\tilde u_{31}\\
      -\tilde u_{12} &             0 &  \tilde u_{23} \\
       \tilde u_{31} & -\tilde u_{23} &             0
     \end{array}
    \right) ,
    \quad
    \tilde V =
  \left(
    \begin{array}{ccc}
                0 &  \tilde v_{12} & -\tilde v_{31}\\
    -\tilde v_{12} &  0            &  \tilde v_{23} \\
     \tilde v_{31} & -\tilde v_{23} &             0
    \end{array}
    \right),
  \quad
\end{equation}
and
\begin{equation}\label{eq:K}
  H = U^T\,dA\,V = 
  \left(
    \begin{array}{ccc}
      k_{11} & k_{12} & k_{13}\\
      k_{21} & k_{22} & k_{23} \\
      k_{31} & k_{32} & k_{33}
  \end{array}
    \right) = dS + \tilde U S + S \tilde V^T,
\end{equation}
we see that in Eq.~(\ref{eq:K}) the diagonal part consists of $dS$, and the nondiagonal part consists of
$\tilde U S + S \tilde V^T$.  Thus $dS$ is made by the diagonal of $H$,
\begin{equation}
  dS = \left(
    \begin{array}{ccc}
    k_{11} &  0 & 0 \\
    0      & k_{22} & 0 \\
    0 & 0 & k_{33}
  \end{array}
    \right).
\end{equation}
For the nondiagonal part, we obtain three linear systems of two equations,
\begin{equation}\label{eq:linsys123}
  \left(
    \begin{array}{c}
    k_{12} \\ k_{21} 
  \end{array} \right)
  =
  \left(
    \begin{array}{cc}
    s_2  & -s_1   \\
    -s_1 &  s_2  
  \end{array} \right)
  \left(
    \begin{array}{c}
    \tilde u_{12} \\ \tilde v_{12}
  \end{array} \right)
,  \qquad \quad
  \left(
    \begin{array}{c}
    k_{23} \\ k_{32} 
  \end{array} \right)
  =
  \left(
    \begin{array}{cc}
    s_3 & -s_2 \\
   -s_2 &  s_3  
  \end{array} \right)
  \left(
    \begin{array}{c}
    \tilde u_{23} \\ \tilde v_{23} 
  \end{array} \right)
, \qquad \quad
  \left(
    \begin{array}{c}
    k_{31}\\ k_{13} 
  \end{array} \right)
  =
  \left(
    \begin{array}{cc}
    s_1 & -s_3  \\
   -s_3 &  s_1  
  \end{array} \right)
  \left(
    \begin{array}{c}
    \tilde u_{31}\\ \tilde v_{31}
  \end{array} \right)
.  
\end{equation}
When we solve these systems for nondegenerate cases $s_1 > s_2 > s_3$, the explicit solution reads\begin{equation}\label{eq:sollinsys123}
  \left(
    \begin{array}{c}
    \tilde u_{12} \\ \tilde v_{12}
    \end{array} \right)
  = 
  { 1\over s_2^2-s_1^2} 
  \left(
    \begin{array}{cc}
      s_2  & s_1   \\
      s_1 &  s_2  
    \end{array} \right)
  \left(
    \begin{array}{c}
      k_{12} \\ k_{21} 
    \end{array} \right)
  ,  \quad
  \left(
    \begin{array}{c}
      \tilde u_{23} \\ \tilde v_{23}
    \end{array} \right)
  = 
  { 1\over s_3^2-s_2^2} 
  \left(
    \begin{array}{cc}
    s_3  & s_2   \\
    s_2 &  s_3  
    \end{array} \right)
  \left(
    \begin{array}{c}
      k_{23} \\ k_{32} 
    \end{array} \right)
  , \quad 
  \left(
    \begin{array}{c}
      \tilde u_{31} \\ \tilde v_{31}
    \end{array} \right)
  = 
  { 1\over s_1^2-s_3^2} 
  \left(
    \begin{array}{cc}
    s_1  & s_3   \\
    s_3 &  s_1  
    \end{array} \right)
  \left(
    \begin{array}{c}
    k_{31} \\ k_{13} 
    \end{array} \right)
  .
\end{equation}
In case of degenerate cases, we can always solve Eqs.\ (\ref{eq:linsys123}) using the pseudo-inverse instead, yielding the
least-norm solution.  Thus we can always obtain the matrices $\tilde U$ and $\tilde V$.

When $A$ is symmetric, $U = V$, $\tilde U = \tilde V$, and $H$ is symmetric as well, so that Eqs.(\ref{eq:sollinsys123}) reduce to
\begin{equation}
  \tilde u_{12}  =  { k_{12}\over s_2-s_1}  , \quad \tilde u_{23}  =  { k_{23}\over s_3-s_2} , \quad \tilde u_{31}  =  { k_{31} \over s_1-s_3} ,
\end{equation}
yielding the nonzero components of $\tilde U$.

Finally, because $\tilde U = U^T dU$, and $U^{-1} = U^T$ (and similar for $\tilde V$), with a last matrix product, the differential
matrices $dU$ and $dV$ are obtained,
\begin{equation}
  dU = U \,\tilde U, \qquad dV= V \,\tilde V.
\end{equation}


\section*{Appendix B: Class of distribution functions with distinct space and velocity pattern speeds}

Here we confirm with a particular class of functions that distribution functions exist that have different pattern speeds, $\Omega$
in space and $\omega$ in velocity space.  We adopt a product of ellipsoidal distribution functions $F$ and $G$ in space and velocity
space, respectively,
\begin{equation}
  f(\vec x, \vec v, t) = F\left(a x(t)^2 + b y(t)^2 + c z^2\right) \, G\left(\alpha v_x(t)^2 + \beta v_y(t)^2 +\gamma  v_z^2\right),  
\end{equation}
where $a$, $b$, $c$, $\alpha$, $\beta$, and $\gamma$ are constants, and
\begin{eqnarray}
  \begin{array}{cc}
    x(t) ~=~ r \cos(\Omega t), & \quad v_x(t) ~=~ v \cos(\omega t), \\
    y(t) ~=~ r \sin(\Omega t), & \quad v_y(t) ~=~ v \sin(\omega t),
  \end{array}
\end{eqnarray}
where $r$ and $v$ are constants.  Clearly, the distribution function rotates at distinct speeds $\Omega$ and $\omega$ in space and
velocity space around the $z-$ and $v_z$-axes, respectively. Direct evaluation of both sides of Eq.\ (\ref{equ:infrot2}) verifies that
the two sides read the same:
\begin{equation}
  -2 \Omega (a-b)  x y F^\prime \left(a x^2+b y^2+c z^2\right)\, G\left(\alpha v_x^2 +\beta v_y^2 +\gamma v_z^2\right)
  -2 \omega (\alpha-\beta)  v_x v_y  F\left(a x^2+b y^2+c z^2\right) \, G^\prime\left(\alpha v_x^2 +\beta v_y^2 +\gamma v_z^2\right) ,
\end{equation}
where $F^\prime$ and $G^\prime$ are the derivatives of $F$ and $G$. The only constraint on $F$ and $G$ therefore is that they should be
differentiable.

\end{document}